\documentclass[aps,prl,twocolumn,reprint]{revtex4}
\usepackage{amsfonts, amssymb, amsmath,graphicx}
\usepackage{subfigure}
\usepackage{multirow}
\usepackage{dcolumn}
\usepackage{bm}
\usepackage{color}

\begin{document}

\title{Anomalous and Quantum Hall Effects in Lossy Photonic Lattices}

\author{Tomoki Ozawa}
\author{Iacopo Carusotto}
\affiliation{
INO-CNR BEC Center and Dipartimento di Fisica, Universit\`a di Trento, I-38123 Povo, Italy
}%

\date{\today}

\def\del{\partial}
\def\p{\prime}
\def\simge{\mathrel{%
         \rlap{\raise 0.511ex \hbox{$>$}}{\lower 0.511ex \hbox{$\sim$}}}}
\def\simle{\mathrel{
         \rlap{\raise 0.511ex \hbox{$<$}}{\lower 0.511ex \hbox{$\sim$}}}}
\newcommand{\feynslash}[1]{{#1\kern-.5em /}}
\newcommand{\iac}[1]{{\color{red} #1} }

\begin{abstract}
We theoretically discuss analogues of the anomalous and the integer quantum Hall effect in driven-dissipative two-dimensional photonic lattices in the presence of a synthetic gauge field. Photons are coherently injected by a spatially localized pump, and the transverse shift of the in-plane light distribution under the effect of an additional uniform force is considered. Depending on pumping parameters, the transverse shift turns out to be proportional either to the global Chern number (integer quantum Hall effect) or to the local Berry curvature (anomalous Hall effect). This suggests a viable route to experimentally measure these quantities in photonic lattices.
\end{abstract}

\maketitle

The amazing developments in the experimental study of quantum fluids of light in the past decade are opening the way to use photonic systems to improve our understanding of phenomena originally known in the context of condensed matter physics~\cite{Carusotto2013}. After pioneering studies of Bose-Einstein condensation~\cite{Kasprzak2006} and superfluidity effects~\cite{Amo2009}, a great interest is presently devoted to topological effects, such as synthetic gauge fields for photons and edge states in photonic topological insulators~\cite{Haldane2008, Wang2009, Rechtsman2013a,Rechtsman2013b,Hafezi2013b,Simon2013}. Inspired by related developments in solid state physics~\cite{Hasan2010, Qi2011, Bernevig2013,Stern2008}, these advances are opening exciting perspectives in the direction of quantum Hall effects with light~\cite{Cho2008, Umucalilar2012, Hafezi2013a} as well as promising applications to photonic devices~\cite{Hafezi2011,Fang2012}.

In this perspective, it is natural to wonder how robust topological effects are against photon losses and the consequent need for an external optical pumping. This question is even more intriguing as Laughlin's gedanken experiment in Ref.~\cite{Laughlin1981} has related the integer quantum Hall effect to gauge invariance, while gauge-dependent quantities such as the photon phase are experimentally accessible in optics, especially under a coherent pumping. 

Crucial concepts in the theoretical description of quantum mechanical particles in periodic lattices under a strong gauge field are the Berry curvature of a band and its integral over the Brillouin zone, the Chern number. This latter is a topological invariant of a band, and, in two-dimensional solid state systems, it is related to the quantized Hall conductance and to the number of chiral edge states~\cite{Thouless1982, Hatsugai1993}. Pioneering experimental studies of these concepts in the photonic context were reported in Refs.~\cite{Wang2009, Rechtsman2013a,Rechtsman2013b,Hafezi2013b,Simon2013}. On the other hand, the local Berry curvature is a geometrical property of a band, which affects various electronic transport properties, in particular, the so-called anomalous Hall conductivity~\cite{Chang1996, Nagaosa2010,Xiao2010}. In the past years, many proposals have appeared to measure it in cold atomic gases trapped in optical lattices~\cite{Dudarev2004, Pettini2011, Price2012, Abanin2013, Alba2011,Goldman2013}. 

In this Letter, we propose a scheme to observe optical analogues of the anomalous and the (integer) quantum Hall effects by using a class of topological photonic devices of high experimental interest, namely, coupled cavity arrays~\cite{Hafezi2013b,AmoPC}. In contrast to the conservative photon dynamics studied in the waveguide experiments of Refs.~\cite{Rechtsman2013a,Rechtsman2013b} and considered in the proposal~\cite{Cominotti2013}, the present work takes advantage of the driven-dissipative nature of the system to relate the Berry curvature and the Chern number to observable quantities. Our ideas are first illustrated on the simplest case of the square-lattice photonic Hofstadter model of~\cite{Hafezi2013b}, and then we generalize the proposal to photonic honeycomb lattices~\cite{AmoPC}, where a nonzero Berry curvature appears in the vicinity of the (gapped) Dirac points when a lattice asymmetry is introduced~\cite{Xiao2007, CastroNeto2009}.

{\it Model.---}
We describe the conservative dynamics of the two-dimensional photonic lattice by using a tight-binding Hamiltonian, which in the square lattice case has the form
\begin{align}
	H
	&=
	\sum_{m,n}
	\left[
	Fn a_{m,n}^\dagger \hat{a}_{m,n}
	-J \left(
	\hat{a}_{m,n}^\dagger \hat{a}_{m+1,n}
	+ \hat{a}_{m+1,n}^\dagger \hat{a}_{m,n}
	\right.
	\right.
	\notag \\
	&\left.\left.
	+ e^{- i 2\pi \alpha m} \hat{a}_{m,n}^\dagger \hat{a}_{m,n+1}
	+ e^{i 2\pi \alpha m} \hat{a}_{m,n+1}^\dagger \hat{a}_{m,n}
	\right)
	\right].
\label{eq:Ham}
\end{align}
Here $\hat{a}_{m,n}$ ($\hat{a}^\dagger_{m,n}$) is the annihilation (creation) operator of a photon on the $(m,n)$ site of the lattice, and the energy zero is set at the energy of the bare cavities.
The hopping along the $\pm x$ direction has a real amplitude $J$, while hopping along the $\pm y$ direction carries an $x$-dependent phase ${\pm 2\pi \alpha m}$, which encodes the synthetic magnetic field acting on the photons, corresponding to the Landau gauge vector potential $\vec{A} = (0, 2\pi \alpha x, 0)$ with a synthetic magnetic flux per lattice plaquette of $\alpha$ in units of the unit magnetic flux. Lengths are measured in units of the lattice spacing.

The first term in (\ref{eq:Ham}) models an external constant force of magnitude $F$ acting on the photons along the $-y$ direction. In the absence of this term, the single-particle physics reduces to the one of charged electrons moving on a square lattice with a perpendicular magnetic field as first considered by Harper and Hofstadter~\cite{Harper1955, Hofstadter1976} in the context of solid state physics. In particular, the energy spectrum $\mathcal{E}$ as a function of $\alpha$ shows a fractal structure known as Hofstadter's butterfly. In the following, we shall assume that the magnetic flux has a rational $\alpha = p/q$ value with coprime integers $p$ and $q$; in this case, we have $q$  energy bands of dispersion $\mathcal{E}_{i}(\mathbf{k})$, whose nontrivial topology is apparent as the local Berry curvature $\Omega_i(\mathbf{k})$ and the global Chern number $2\pi \mathcal{C}_i=\int_{\mathrm{MBZ}} d^2\mathbf{k}\,\Omega_i(\mathbf{k})$ are nonzero for each of them, where the last integral is over the magnetic Brillouin zone (MBZ) defined by $[-\pi/q,\pi/q]\times[-\pi,\pi]$.

As we are considering a driven-dissipative photonic lattice, we have to include the effect of pumping and losses~\cite{Carusotto2013}. Losses are assumed to be local and uniform for all lattice sites at a rate $\gamma$. The pumping field is taken to be monochromatic with frequency $\omega_0$ and a spatial amplitude profile $f_{m,n}$.
In the linear optics case under consideration here, photons are noninteracting so exact results are obtained by the mean-field equations for the expectation values $a_{m,n}(t) =\langle \hat{a}_{m,n}(t) \rangle$. In the steady state, these evolve according to the harmonic law $a_{m,n}(t) = a_{m,n} e^{-i\omega_0 t}$ with time-independent amplitudes $a_{m,n}$ satisfying the linear system
\begin{align}
	&J\left[
	a_{m+1,n} + a_{m-1,n}
	+ e^{-i2\pi \alpha m} a_{m,n+1} + e^{i2\pi \alpha m} a_{m,n-1}
	\right]
	\notag \\
	&\hspace{2.5cm}+\left[
	\omega_0 + i\gamma - Fn
	\right]
	a_{m,n}
	=
	f_{m,n}. \label{steadyeq}
\end{align}
which can be numerically solved on a finite lattice.
In the following, we shall assume that only the central site $(0,0)$ is pumped: $f_{m,n} = f \delta_{m,0}\delta_{n,0}$.

\begin{figure}[htbp]
\begin{center}
\includegraphics[width=7.8cm]{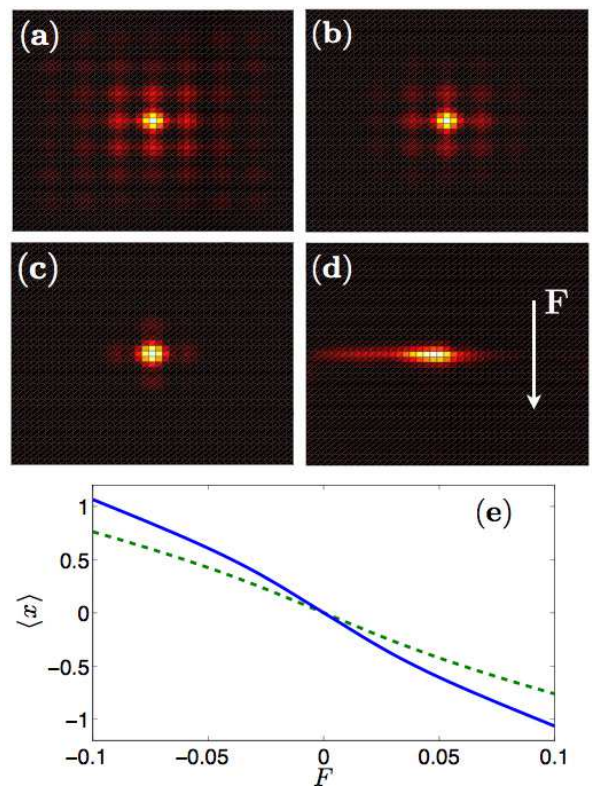}
\caption{(a)-(d) Photon amplitude distribution $|a_{m,n}|$ on a $41 \times 41$ square lattice with $\alpha = 1/5$.
The central sites are pumped.
The force $F$ is zero for (a)-(c), and $F = 0.1J$ for (d). In (a),(b),(d), the pump frequency is tuned to $\omega_0/J = -2.95$ within the lowest energy band; in (c), it is tuned to $\omega_0/J = -2.85$ within a band gap. The loss rate is $\gamma = 0.01J$ for (a),(c),(d) and $\gamma = 0.02J$ for (b). The bright regions have higher intensity than the dark regions.
(e) Displacement $\langle x \rangle$ as a function of $F$, in units of $J$, for a pump frequency $\omega_0/J = -2.95$ with $\alpha = 1/5$. The solid (blue) line is for $\gamma/J = 0.05$, and the dashed (green) line is for $\gamma/J = 0.08$.}
\label{intensity_extent}
\end{center}
\end{figure}

This physics is illustrated in Fig.~\ref{intensity_extent} starting from the $F=0$ case with no external force: In Figs.~\ref{intensity_extent}(a) and \ref{intensity_extent}(b), the pump frequency is chosen within the lowest magnetic band of $\alpha = 1/5$. As the loss rate $\gamma$ is increased from $\gamma = 0.01J$ (a) to $\gamma = 0.02J$ (b), photons are able to travel over shorter distances before decaying, so the photon intensity distribution gets more and more spatially localized in the vicinity of the pumped site: Rather than a hindrance, the lossy nature of the system is here a useful tool to suppress spurious effects due to the lattice edges.
The exponential localization effect is even more dramatic when the frequency falls within a band gap [Fig.~\ref{intensity_extent}(c)], and the bands are excited in a nonresonant way.

{\it Measuring topological quantities.---}
The situation becomes more interesting once we turn on the synthetic electric field $F \neq 0$ directed along the negative $y$ direction: From Fig.~\ref{intensity_extent}(d), it is apparent that the photon intensity distribution is no longer centered at the pump position but is significantly shifted in the leftward direction transverse to the applied force.
Examples of the dependence of the transverse displacement of the center of mass $\langle x \rangle \equiv [\sum_{m,n} m |a_{m,n}|^2 ]/ [\sum_{m,n} |a_{m,n}|^2 ]$ on the applied force $F$ are displayed
in Fig.~\ref{intensity_extent}(e), where we plot $\langle x\rangle$ as a function of $F$ for a pump frequency within the lowest energy band of $\alpha = 1/5$ and two different loss values $\gamma/J = 0.05$ and $0.08$. The displacement $\langle x \rangle$ grows linearly for small $F$; for the parameters in the figure, this linear regime extends up to $|F| \simle 0.02 J$. 

We now proceed to relate the slope of this linear dependence to the topological properties of the band; a single band description is legitimate, provided the pump frequency $\omega_0$ falls within (or close to) an energy band and $\gamma$ is smaller than the band gap separating from the next bands. In the linear regime, this gives the simple relation between the displacement and the Berry curvature (a full proof of (\ref{meanxresult})  as well as its extension to more complex - e.g., honeycomb - lattices is given in  Supplemental Material)
\begin{align}
	\langle x \rangle
	=
	F
	\frac{
	\int_{\mathrm{MBZ}} \gamma\,\Omega (\mathbf{k})\,n(\mathbf{k})^2
	}{
	\int_{\mathrm{MBZ}} n(\mathbf{k})
	}, \label{meanxresult}
\end{align}
where $n(\mathbf{k})=[(\omega_0-\mathcal{E}(\mathbf{k}))^2+\gamma^2]^{-1}$ is the (normalized) population distribution within the band under consideration; $\mathcal{E}(\mathbf{k})$ and $\Omega (\mathbf{k})$ are the energy dispersion and the local Berry curvature, respectively, of the corresponding band.

The integral in (\ref{meanxresult}) can be worked out explicitly in the two cases of large and small $\gamma$ as compared to the bandwidth of the energy band $\Delta_{\mathrm{width}}$; for large $\gamma\gg\Delta_{\mathrm{width}}$, one obtains the Chern numbers, and for small $\gamma\ll\Delta_{\mathrm{width}}$, one finds the Berry curvature. In practical experiments, the loss rate $\gamma$ can be tuned by artificially reducing the quality factor of the cavities or, alternatively, by tuning $\Delta_{\mathrm{width}}$ by varying the hopping amplitude $J$.

{\it Large loss: Chern number and quantum Hall effect.---}
In the limit $\gamma \gg \Delta_{\mathrm{width}}$, the detuning term $[\omega_0-\mathcal{E}(\mathbf{k})]^2$ in $n(\mathbf{k})$ can be neglected, and a formula for the transverse shift is found:
\begin{align}
	\langle x \rangle
	&\approx
	F
	\dfrac{
	\int_{\mathrm{MBZ}}d^2 \mathbf{k}\,
	\Omega (\mathbf{k}) / \gamma^3
	}{
	\int_{\mathrm{MBZ}}d^2 \mathbf{k}\,
	1 / \gamma^2
	}
	= \frac{q \mathcal{C} F}{2\pi \gamma} \label{xchern}
\end{align}
that involves only the Chern number $\mathcal{C}$ of the band: This result is an optical analogue of the integer quantum Hall effect. 

Of course, this formula is valid only if the loss rate $\gamma$ is much smaller than the band gap to the nearest energy band $\Delta_{\mathrm{gap}}$. 
This condition imposes a compromise between a large enough value of $\gamma/J$ to encompass the whole band of interest and a small enough value to minimize the spurious effect of the neighboring bands. For the lowest band of $\alpha=1/5$, the large separation from the higher bands ($\Delta_{\mathrm{gap}} / \Delta_{\mathrm{width}} \sim 24$) allows for a good compromise: By using $\gamma= 2\Delta_{\mathrm{width}}$ and $\omega_0$ tuned at the band center, the estimated value $\mathcal{C}_n \approx -0.96$ of the Chern number is close to the exact value $\mathcal{C}=-1$.

As the leading order correction due to the neighboring bands to (\ref{xchern}) does not depend on $\gamma$, a more precise estimate can be obtained by repeating the measurement on different samples with different values of the normalized loss rate $\gamma/J$ so as to extract the coefficient of $1/\gamma$ in (\ref{xchern}). In Table~\ref{cherntable}, we list the estimated Chern numbers $\mathcal{C}_n$ obtained by numerically calculating the mean displacement $\langle x\rangle$ for different values of $\alpha$, and we compare them with the exact values $\mathcal{C}$ obtained from the Diophantine equation approach~\cite{Bernevig2013, Thouless1982}. For each case, the pump frequency $\omega_0$ is chosen to be at the center of the band under examination, and the coefficient of the $1/\gamma$ term is calculated for a normalized loss value $\gamma/2J = (J/\Delta_{\mathrm{width}} + J/\Delta_{\mathrm{gap}})^{-1}$.

As long as the bandwidth is much larger than the band gap, the agreement is very good. 
On the other hand, the large deviation between the estimated and the exact Chern numbers of the second and fourth bands of $\alpha = 1/5$ ($\mathcal{C}_n = -0.66$ instead of $\mathcal{C}=-1$ ) is because the corresponding bandwidth $\Delta_{\mathrm{width}} \sim 0.45J$ is very close to the size of the band gap $\Delta_{\mathrm{gap}} \sim 0.52J$. When $\Delta_{\mathrm{width}} > \Delta_{\mathrm{gap}}$, this method is not reliable at all, so the corresponding cases in the table have been left blank.

\begin{table}[htb]
\begin{tabular}{|l|c|c|c|c|c|c|c|} \hline
 &  & 1st & 2nd & 3rd & 4th & 5th & 6th \\ \hline \hline
\multirow{2}{*}{$\alpha = \dfrac{1}{3}$} & $\mathcal{C}$ & $-1$ & $+2$ & $-1$ \\ \cline{2-5}
& $\mathcal{C}_n$ & $-0.91$ & - & $-0.91$ \\ \cline{1-7}
\multirow{2}{*}{$\alpha = \dfrac{1}{5}$} & $\mathcal{C}$ & $-1$ & $-1$ & $+4$ & $-1$ & $-1$ \\ \cline{2-7}
& $\mathcal{C}_n$ & $-0.97$ & $-0.66^{*}$ & - & $-0.66^{*}$ & $-0.97$ \\ \cline{1-8}
\multirow{2}{*}{$\alpha = \dfrac{1}{6}$} & $\mathcal{C}$ & $-1$ & $-1$ & $+2$ & $+2$ & $-1$ & $ -1$  \\ \cline{2-8}
& $\mathcal{C}_n$ & $-0.96$ & $-1.06$ & - & - & $-1.06$ & $-0.96$ \\ \cline{1-8}
\multirow{2}{*}{$\alpha = \dfrac{3}{7}$} & $\mathcal{C}$ & $+2$ & $-5$ & $+2$ & $+2$ & $+2$ & $-5$ \\ \cline{2-8}
& $\mathcal{C}_n$ & $2.05$ & - & - & $2.01$ & - & - \\ \cline{1-8}
\multirow{2}{*}{$\alpha = \dfrac{4}{9}$} & $\mathcal{C}$ & $+2$ & $+2$ & $-7$ & $+2$ & $+2$ & $+2$ \\ \cline{2-8}
& $\mathcal{C}_n$ & $1.96$ & - & - & $2.02$ & $1.92$ & $2.02$ \\ \hline
\end{tabular}
\caption{Table of the estimated Chern numbers $\mathcal{C}_n$ of the photonic bands, compared to the real Chern numbers $\mathcal{C}$, for several values of $\alpha$. The numerical estimations are obtained by implementing the method discussed in the text on $41 \times 41$ square lattices. The blank cases indicate bands for which $ \Delta_{\mathrm{width}}<  \Delta_{\mathrm{gap}}$ where the method is not reliable; the $*$ signs indicate bands for which $ \Delta_{\mathrm{width}}\simeq \Delta_{\mathrm{gap}}$ and large discrepancies are expected.}
\label{cherntable}
\end{table}

{\it Small loss: Berry curvature and anomalous Hall effect.---}
When the loss $\gamma$ is much smaller than the bandwidth $\Delta_{\mathrm{width}}$, the $\mathbf{k}$-space distribution can be approximated as a delta function on the $\mathbf{k}$-space locus where $\omega_0 \approx \mathcal{E}(\mathbf{k})$. In analogy to the (intrinsic contribution to the) anomalous Hall effect, the displacement
\begin{align}
	\langle x \rangle
	&\approx
	F \gamma \bar{\Omega}(\omega_0)
	\dfrac{
	\int_{\mathrm{MBZ}}d^2 \mathbf{k}\,
	n(\mathbf{k})^2
	}{
	\int_{\mathrm{MBZ}}d^2 \mathbf{k}\,
	n(\mathbf{k})
	}
	\approx
	\frac{\bar{\Omega}(\omega_0)\,F}{2\gamma} \label{berryfromx}
\end{align}
turns out to be proportional to the average $\bar{\Omega}(\omega_0)$ of the  Berry curvature  on the $\mathcal{E}(\mathbf{k}) = \omega_0$ curve in $\mathbf{k}$ space. Remarkably, different regions of the Brillouin zone can be separately addressed just by tuning the frequency $\omega_0$ of the coherent pump. In the case of the Hofstadter lattice, numerical calculations suggest that the Berry curvature is a function of the energy only, so a reliable estimate of the local Berry curvature at the different points of the MBZ can be obtained from a measurement of $\bar{\Omega}(\omega_0)$ provided only the iso-$\mathcal{E}$ locus in $\mathbf{k}$ space does not cross stationary points of $\mathcal{E}(\mathbf{k})$.

\begin{figure}[htbp]
\begin{center}
\includegraphics[width=8.4cm]{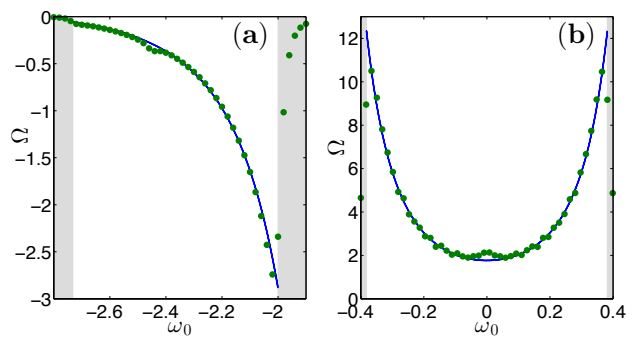}
\caption{The estimated Berry curvature $\Omega$ as a function of $\omega_0$, in units of $J$ (dots) as compared to the exact value (solid lines) for $\gamma/\Delta_{\mathrm{width}}=1/30$ on $201 \times 201$ lattices. (a) Lowest band of square lattice at $\alpha = 1/3$ and (b) middle band of square lattice at $\alpha = 1/5$.}
\label{curvature_combined}
\end{center}
\end{figure}

The accuracy of this result is validated in Fig.~\ref{curvature_combined}, where we plot the estimated value of the Berry curvature for the lowest band of $\alpha = 1/3$ and the central band of $\alpha = 1/5$, and we compare them with the true values. The loss rate is taken here to be $\gamma/\Delta_{\mathrm{width}}=1/30$; for such a small value of $\gamma$, photons propagate over longer distances before decaying, so larger lattices are needed to suppress the effect of the edges. Overall, the estimated value of the Berry curvature agrees well with its exact value for almost all values of the pump frequency $\omega_0$ in the bulk of the bands. Of course, the method breaks down in the vicinity of the band edges and becomes meaningless within the energy gaps (indicated by the gray shading in the figure). But note also the small spurious bumps around $\omega_0/J\sim -2.45$ for $\alpha = 1/3$ and $\omega_0/J \sim 0$ for $\alpha = 1/5$: These deviations correspond to stationary points of the band structure $\mathcal{E}(\mathbf{k})$ where the last equality of (\ref{berryfromx}) is no longer valid. The quantitative discrepancy gets suppressed if smaller values of $\gamma$ are used.
It is, however, important to notice that, since our scheme does not benefit from topological protection, the actual measured displacement $\langle x \rangle$ can be affected by disorder. To suppress the deleterious effect of disorder, one may repeat the measurement by choosing different lattice sites for the pumping and then taking an average.

\begin{figure}[htbp]
\begin{center}
\includegraphics[width=8.5cm]{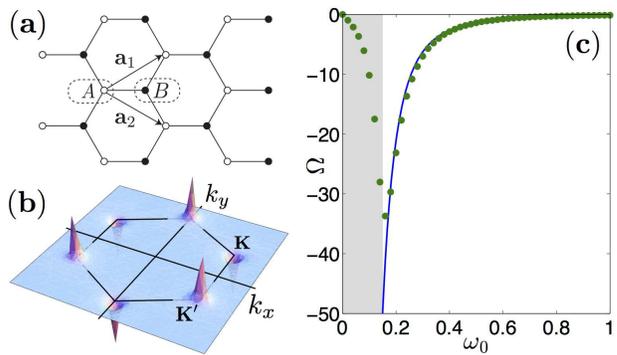}
\caption{(a) Photonic honeycomb lattice. Sites belonging to the sublattices $A$ and $B$ are denoted by empty and filled circles, respectively.
(b) The Berry curvature of the upper band of a honeycomb lattice with $\Delta = 0.3J$. The Dirac points are the vertices of the hexagon.
(c) The estimated Berry curvature $\Omega$ as a function of $\omega_0$, in units of $J$ (dots) as compared to the exact value (solid line), for the honeycomb lattice without a synthetic magnetic field. System parameters are $\gamma/J = 0.04$ and $\sigma = 3$ on a $200 \times 200$ unit cell honeycomb lattice with site energy mismatch $\Delta/J = 0.3$. The shaded region indicates the band gap.}
\label{graphene_combined}
\end{center}
\end{figure}

{\it Photonic honeycomb lattice.---}
As a last point, we discuss the nontrivial new features that arise when extending our study to honeycomb lattices.
We consider the usual tight-binding model of the honeycomb lattice sketched in Fig.~\ref{graphene_combined}(a), with a nearest-neighbor hopping $J$ which is now real and equal for all links~\cite{CastroNeto2009}.
The unit vectors are $\mathbf{a}_1 \equiv (3/2, \sqrt{3}/2)$ and $\mathbf{a}_2 \equiv (3/2, -\sqrt{3}/2)$, where the distance between two sites is taken to be unity.
In the presence of a small energy difference $\Delta$ between the sublattices, the band degeneracy at the Dirac points $\mathbf{K,~K'} = 2\pi(1/3, \pm1/3\sqrt{3})$ is lifted by a band gap $\Delta$, and the two bands have a nontrivial Berry curvature even in the absence of a synthetic magnetic field~\cite{Xiao2007}; as illustrated in Fig.~\ref{graphene_combined}(b), the Berry curvature is concentrated in the vicinity of the (gapped) Dirac points, and, for each band, it is approximately a function of the energy only. As a result of the time-reversal symmetry, the Berry curvatures at $\mathbf{K}$ and $\mathbf{K}^\prime$ exactly compensate giving a vanishing global Chern number for both bands and therefore no quantum Hall effect.
As the averaged Berry curvature vanishes on any isoenergy curve, extension of the analogue anomalous Hall effect to honeycomb geometries requires separating the contributions of the $\mathbf{K}$ and $\mathbf{K}^\prime$ points.

The simplest strategy in this direction is to use a spatially extended pump with a finite in-plane wave vector in the vicinity of, e.g., the $\mathbf{K}$ point:
\begin{align}
	f({\mathbf{R}}) = f\,e^{-{R}^2/2\sigma^2} e^{i\mathbf{K}\cdot \mathbf{R}}, \label{pumpgraphene}
\end{align}
where $\sigma$ is the spatial extent of the pump spot and $\mathbf{R}$ is the position of each site. In the experimental setup of Ref.~\cite{AmoPC}, this can be obtained by shining a laser field on the microcavity at a finite and well-chosen angle with the microcavity axis~\cite{Carusotto2013}.
As before, the basic idea is to find out how the Berry curvature controls the transverse displacement of the intensity distribution when an in-plane force is applied to the photons.

While doing this, one has to deal with further complications stemming from the inequivalence of the two sublattices: As a result of this, measurements using a pump in the form (\ref{pumpgraphene}) would in fact lead to nonzero displacements even for $F=0$ and to a slope not directly related to the Berry curvature.
All these difficulties can be resolved by repeating the measurement using two different, symmetric frequencies $\pm\omega_0$ and taking a weighted difference of the two measured displacements:
\begin{align}
	\langle x \rangle_-
	&\equiv
	\frac{\sum_{\mathbf{R}}R_x (|a_\mathbf{R}|^2 - |a^\prime_\mathbf{R}|^2)}
	{\sum_{\mathbf{R}}(|a_\mathbf{R}|^2 + |a^\prime_\mathbf{R}|^2)}, \label{meanxgraphene}
\end{align}
where $a_\mathbf{R}$ is the photon field at position $\mathbf{R}$ when the pump frequency is $\omega_0$ and $a^\prime_\mathbf{R}$ is when the frequency is $-\omega_0$.
With this definition of $\langle x \rangle_-$, a straightforward analytical calculation (see the  Supplemental Material for details and also for an alternative strategy to measure the Berry curvature right at the tip of the gapped Dirac cone) shows that the formula (\ref{meanxresult}) holds with the integral taken over the Brillouin zone and the integrand being multiplied by the Fourier transform of the source. Then, using a source which uniformly covers the vicinity of $\mathbf{K}$ but is very small near other Dirac cones, one can estimate the Berry curvature from the transverse shift using the anomalous Hall effect formula (\ref{berryfromx}).
The accuracy of this method is validated in Fig.~\ref{graphene_combined}(c), where we plot the estimated value of the Berry curvature using this method for the region around the $\mathbf{K}$ Dirac point; agreement with the theoretical value (solid line) is very good.

{\it Conclusion.---}
In this Letter, we have discussed optical analogues of the anomalous and integer quantum Hall effects in driven-dissipative photonic lattices with nontrivial geometrical and topological properties. Our results suggest that both the Chern number and the local Berry curvature of the photonic bands can be experimentally extracted from the transverse displacement of the light distribution under the effect of an additional force
and support the promise of photonic cavity arrays as a platform to study the interplay of the nontrivial band topology with many-body physics.

\begin{acknowledgements}
We are grateful to A. Amo for continuous exchanges on photonic honeycomb lattices and to M. Hafezi, M. Rechtsman, Y. Plotnik, and R. O. Umucal\i lar for discussions on topological photonic systems. These scientific exchanges were supported by the POLATOM ESF network.
This work was partially funded by ERC through the QGBE grant and by Provincia Autonoma di Trento.
\end{acknowledgements}


\begin{thebibliography}{99}

\bibitem{Carusotto2013} I. Carusotto and C. Ciuti, Rev. Mod. Phys. {\bf 85}, 299 (2013).

\bibitem{Kasprzak2006} J. Kasprzak, M. Richard, S. Kundermann, A. Baas, P. Jeambrun, J. M. J. Keeling, F. M. Marchetti, M. H. Szyma\'nska, R. Andr\'e, J. L. Staehli, V. Savona, P. B. Littlewood, B. Deveaud, and L. S. Dang, Nature (London) {\bf 443}, 409 (2006).

\bibitem{Amo2009} A. Amo, J. Lefr\`ere, S. Pigeon, C. Adrados, C. Ciuti, I. Carusotto, R. Houdr\'e, E. Giacobino, and A. Bramati, Nature Physics {\bf 5}, 805 (2009).

\bibitem{Haldane2008} F. D. M. Haldane and S. Raghu, Phys. Rev. Lett. {\bf 100}, 013904 (2008).

\bibitem{Wang2009} Z. Wang, Y. Chong, J. D. Joannopoulos, and M. Solja\v{c}i\'{c}, Nature (London) {\bf 461}, 772 (2009).

\bibitem{Rechtsman2013b} M. C. Rechtsman, J. M. Zeuner, Y. Plotnik, Y. Lumer, D. Podolsky, F. Dreisow, S. Nolte, M. Segev, and A. Szameit, Nature {\bf 496}, 196 (2013).

\bibitem{Rechtsman2013a} M. C. Rechtsman, J. M. Zeuner, A. T\"unnermann, S. Nolte, M. Segev, and A. Szameit, Nature Photonics {\bf 7}, 153 (2013).

\bibitem{Hafezi2013b} M. Hafezi, J. Fan, A. Migdall, and J. Taylor, Nature Photonics {\bf 7}, 1001 (2013).

\bibitem{Simon2013}  N. Jia, A. Sommer, D. Schuster, and J. Simon, arXiv:1309.0878.

\bibitem{Hasan2010} M. Z. Hasan and C. L. Kane, Rev. Mod. Phys. {\bf 82}, 3045 (2010).

\bibitem{Qi2011} X.-L. Qi and S.-C. Zhang, Rev. Mod. Phys. {\bf 83}, 1057 (2011).

\bibitem{Bernevig2013} B. A. Bernevig and T. L. Hughes, {\it Topological Insulators and Topological Superconductors} (Princeton University, Princeton, NJ, 2013).

\bibitem{Stern2008} A. Stern, Ann. Phys. (N.Y.) {\bf 323}, 204 (2008).

\bibitem{Cho2008} J. Cho, D. G. Angelakis, and S. Bose, Phys. Rev. Lett. {\bf 101}, 246809 (2008).

\bibitem{Umucalilar2012} R. O. Umucal{\i}lar and I. Carusotto, Phys. Rev. Lett. {\bf 108}, 206809 (2012).

\bibitem{Hafezi2013a} M. Hafezi, M. D. Lukin, and J. M. Taylor, New J. Phys. {\bf 15}, 063001 (2013).

\bibitem{Hafezi2011} M. Hafezi, E. A. Demler, M. D. Lukin, and J. M. Taylor, Nature Physics {\bf 7}, 907 (2011).

\bibitem{Fang2012} K. Fang, Z. Yu, and S. Fan, Nature Photonics {\bf 6}, 782 (2012).

\bibitem{Laughlin1981} R. B. Laughlin, Phys. Rev. B {\bf 23}, 5632 (1981).

\bibitem{Thouless1982} D. J. Thouless, M. Kohmoto, M. P. Nightingale, and M. den Nijs, Phys. Rev. Lett. {\bf 49}, 405 (1982).

\bibitem{Hatsugai1993} Y. Hatsugai, Phys. Rev. Lett. {\bf 71}, 3697 (1993).

\bibitem{Xiao2010} D. Xiao, M.-C. Chang, and Q. Niu, Rev. Mod. Phys. {\bf 82}, 1959 (2010).

\bibitem{Chang1996}  M.-C. Chang and Q. Niu, Phys. Rev. B {\bf 53}, 7010 (1996).

\bibitem{Nagaosa2010} N. Nagaosa, J. Sinova, S. Onoda, A. H. MacDonald, and N. P. Ong, Rev. Mod. Phys. {\bf 82}, 1539 (2010).

\bibitem{Dudarev2004} A. M. Dudarev, R. B. Diener, I. Carusotto, and Q. Niu, Phys. Rev. Lett. {\bf 92}, 153005 (2004).

\bibitem{Pettini2011} G. Pettini and M. Modugno, Phys. Rev. A {\bf 83}, 013619 (2011).

\bibitem{Price2012} H. M. Price and N. R. Cooper, Phys. Rev. A {\bf 85}, 033620 (2012).

\bibitem{Abanin2013} D. A. Abanin, T. Kitagawa, I. Bloch, and E. Demler, Phys. Rev. Lett. {\bf 110}, 165304 (2013).

\bibitem{Alba2011} E. Alba, X. Fernandez-Gonzalvo, J. Mur-Petit, J. K. Pachos, and J. J. Garcia-Ripoll, Phys. Rev. Lett. {\bf 107}, 235301 (2011).

\bibitem{Goldman2013} A. Dauphin and N. Goldman, Phys. Rev. Lett. {\bf 111}, 135302 (2013).

\bibitem{Cominotti2013} M. Cominotti and I. Carusotto, Europhys. Lett. {\bf 103} 10001 (2013).

\bibitem{AmoPC} T. Jacqmin, I. Carusotto, I. Sagnes, M. Abbarchi, D. Solnyshkov, G. Malpuech, E. Galopin, A. Lema\^{i}tre, J. Bloch, and A. Amo, arXiv:1310.8105.

\bibitem{Xiao2007} D. Xiao, W. Yao, and Q. Niu, Phys. Rev. Lett. {\bf 99}, 236809 (2007).

\bibitem{CastroNeto2009} A. H. Castro Neto, N. M. R. Peres, K. S. Novoselov, and A. K. Geim, Rev. Mod. Phys. 81, {\bf 109} (2009).

\bibitem{Harper1955} P. G. Harper, Proc. Phys. Soc. London Sect. A {\bf 68}, 874 (1955).

\bibitem{Hofstadter1976} D. R. Hofstadter, Phys. Rev. B {\bf 14}, 2239 (1976).

\end{thebibliography}
\end{document}


\title{Supplemental Material for ``Anomalous and Quantum Hall Effects in Lossy Photonic Lattices"}

\author{Tomoki Ozawa}
\author{Iacopo Carusotto}
\affiliation{
INO-CNR BEC Center and Dipartimento di Fisica, Universit\`a di Trento, I-38123 Povo, Italy
}%

\def\del{\partial}
\def\p{\prime}
\def\simge{\mathrel{%
         \rlap{\raise 0.511ex \hbox{$>$}}{\lower 0.511ex \hbox{$\sim$}}}}
\def\simle{\mathrel{
         \rlap{\raise 0.511ex \hbox{$<$}}{\lower 0.511ex \hbox{$\sim$}}}}
\newcommand{\feynslash}[1]{{#1\kern-.5em /}}
\newcommand{\iac}[1]{{\color{red} #1} }

\maketitle

\appendix

\renewcommand{\theequation}{A\arabic{equation}}
\section{Appendix A: Mean displacement and Berry curvature}

We give an outline of the derivation of Eq.~(3) of the main text, which connects the mean displacement $\langle x \rangle$ and the Berry curvature $\Omega (\mathbf{k})$.
We first Fourier transform the photon fields $a_{m,n}$ to work in the momentum space.
Since the magnetic unit cell contains $q$ original unit cells in $x$ direction, $a_{m,n}$ can be expanded as
\begin{align}
	a_{m,n}
	=
	\sum_{k_x, k_y} e^{ik_x m + ik_y n} c_{m^{\prime \prime}}(k_x, k_y),
\end{align}
where $c_{m^{\prime \prime}}(k_x, k_y)$ are the annihilation operators in momentum space and $m = qm^\prime + m^{\prime \prime}$ with $m^\prime$ being an integer and $m^{\prime \prime} = 1, 2, \cdots, q$. The sum over the wave vectors is restricted to the magnetic Brillouin zone $-\pi/q < k_x < \pi/q$ and $-\pi < k_y < \pi$.
In terms of the photon fields in momentum space, the mean displacement is
\begin{align}
	\langle x \rangle
	=
	\frac{\sum_{k_x, k_y} \langle c(\mathbf{k}) | i\frac{\partial}{\partial k_x}| c(\mathbf{k})\rangle}
	{\sum_{k_x, k_y} \langle c(\mathbf{k}) | c(\mathbf{k})\rangle}, \label{meanxinbloch}
\end{align}
where $|c (\mathbf{k})\rangle \equiv (c_1 (\mathbf{k}), c_2 (\mathbf{k}), \cdots, c_q (\mathbf{k}))$ is the q-component vector.

After a straightforward algebra, the steady-state equation of motion Eq.~(2) can be written in momentum space as
\begin{align}
	\left[
	\left(
	\omega_0 + i\gamma - iF\frac{\partial}{\partial k_y}
	\right)
	I_q
	-
	H_\mathbf{k}
	\right]
	|c(\mathbf{k})\rangle
	=
	|f_q\rangle, \label{blocheom}
\end{align}
where $I_q$ is a $q \times q$ identity matrix and 
\begin{align}
	&H_\mathbf{k}
	\equiv -2J \times
	\notag \\
	&
	\begin{pmatrix}
	\cos (2\pi \alpha - k_y) & e^{ik_x}/2 & \cdots & e^{-ik_x}/2 \\
	e^{-ik_x}/2 & \cos (4\pi \alpha - k_y) & \cdots & 0 \\
	0 & e^{-ik_x}/2 & \cdots & 0 \\
	\vdots & \vdots & \ddots & \vdots \\
	e^{ik_x}/2 & 0 & \cdots & \cos (2\pi q \alpha - k_y ) \\
	\end{pmatrix}
\end{align}
is the $q \times q$ tight-binding Bloch Hamiltonian of a charged particle in a two-dimensional square lattice with perpendicular magnetic field, and we defined a $q$ component vector
$|f_q\rangle \equiv fq/(L_x L_y) (0, 0, \cdots, 0, 1)$ with $L_x$ and $L_y$ being the linear size of the system in $x$ and $y$ directions, respectively, which is the Fourier transform of the pump field whose only nonzero component is the last term.

The steady-state photon fields in momentum space can be obtained by solving Eq.~(\ref{blocheom}), which can be expanded in a power series of $F$ as
\begin{align}
	&|c(\mathbf{k}) \rangle
	\approx
	\left[(\omega_0 + i\gamma) I_q - H_\mathbf{k}\right]^{-1}|f_q\rangle
	\notag \\
	&
	+
	\left[(\omega_0 + i\gamma) I_q - H_\mathbf{k}\right]^{-1}
	iF\frac{\partial}{\partial k_y}
	\left[(\omega_0 + i\gamma) I_q - H_\mathbf{k}\right]^{-1}
	|f_q\rangle.
\end{align}
The eigenvectors of $\left[(\omega_0 + i\gamma) I_q - H_\mathbf{k}\right]^{-1}$ are the same as the eigenstates of the Bloch Hamiltonian $H_\mathbf{k}$:
\begin{align}
	\left[(\omega_0 + i\gamma) I_q - H_\mathbf{k}\right]^{-1}|u_i (\mathbf{k})\rangle
	=
	\frac{1}{\omega_0 + i\gamma - \mathcal{E}_i (\mathbf{k})}|u_i (\mathbf{k})\rangle,
\end{align}
where $|u_i (\mathbf{k})\rangle$ is the normalized eigenstates corresponding to $i$-th band of $H_\mathbf{k}$ with eigenenergy $\mathcal{E}_i (\mathbf{k})$.
Then, assuming that $\omega_0$ is tuned to one magnetic energy band, which we take to be the first band without loss of generality, we may ignore the effects from other bands and approximate
\begin{align}
	&\left[(\omega_0 + i\gamma) I_q - H_\mathbf{k}\right]^{-1}|f_q\rangle
	\notag \\
	&=
	\sum_i \frac{\langle u_i (\mathbf{k}) | f_q \rangle}{\omega_0 + i\gamma - \mathcal{E}_i (\mathbf{k})}|u_i (\mathbf{k})\rangle
	\approx
	\frac{\langle u_1 (\mathbf{k}) | f_q \rangle}{\omega_0 + i\gamma - \mathcal{E}_1 (\mathbf{k})}|u_1 (\mathbf{k})\rangle.
\end{align}
Then, the denominator of (\ref{meanxinbloch}) is, converting the sum into an integral in the magnetic Brillouin zone,
\begin{align}
	\int_{\mathrm{MBZ}}d^2 \mathbf{k}\, \langle c(\mathbf{k}) | c(\mathbf{k})\rangle
	=
	\int_{\mathrm{MBZ}}d^2 \mathbf{k} \frac{\langle u_1 (\mathbf{k}) | f_q \rangle \langle f_q | u_1 (\mathbf{k})\rangle}{(\omega_0 - \mathcal{E}_1 (\mathbf{k}))^2 + \gamma^2}.
\end{align}
Instead of $|f_q\rangle$, one can use $|f_i\rangle$ for the Fourier transform of the source term and we should obtain the same result, where $|f_i\rangle$ is the $q$-component vector whose magnitude is the same as $|f_q\rangle$, but only the $i$-th component is nonzero.
This is because using $|f_i\rangle$ corresponds to using the site $(i,0)$ for the pump site, which should not change the overall physics as long as the edge effects can be ignored.
Then, using $\sum_i |f_i \rangle \langle f_i| = \langle f_q | f_q \rangle I_q$, we obtain
\begin{align}
	\int_{\mathrm{MBZ}}d^2 \mathbf{k}\,\langle c(\mathbf{k}) | c(\mathbf{k})\rangle
	=
	\int_{\mathrm{MBZ}}d^2 \mathbf{k}\, \frac{\langle f_q | f_q \rangle}{q}n(\mathbf{k}), \label{denresult}
\end{align}
where $n(\mathbf{k}) \equiv [(\omega_0-\mathcal{E}(\mathbf{k})^2+\gamma^2]^{-1}$ is the normalized population distribution within the band, as introduced in the main text.
Similarly, but after a more lengthy algebra, one obtains for the numerator of Eq.~(\ref{meanxinbloch}), up to the first order in $F$,
\begin{align}
	&\int_{\mathrm{MBZ}}d^2 k \langle c(\mathbf{k}) |i\frac{\partial}{\partial k_x}| c(\mathbf{k})\rangle
	=\frac{\langle f_q | f_q \rangle}{q} \times\notag \\
	&
	\int_{\mathrm{MBZ}}d^2 k
	\left\{
	\gamma \frac{\partial \mathcal{E}}{\partial k_x}
	+
	F
	\left(
	\gamma \Omega (\mathbf{k})
	+
	\mathcal{A}_{\mathrm{sym}}(\mathbf{k})
	\right)
	\right\}
	n(\mathbf{k})^2,
	\label{numresult}
\end{align}
where
\begin{align}
	\Omega (\mathbf{k})
	\equiv
	i\left[
	\left\langle \frac{\partial u_1}{\partial k_x} \right|\left. \frac{\partial u_1}{\partial k_y} \right\rangle
	-
	\left\langle \frac{\partial u_1}{\partial k_y} \right|\left. \frac{\partial u_1}{\partial k_x} \right\rangle
	\right],
\end{align}
is the Berry curvature of the band at momentum $\mathbf{k}$, and
$\mathcal{A}_{\mathrm{sym}}(\mathbf{k})$ is a term symmetric under the exchange of $\partial/\partial k_x$ and $\partial/\partial k_y$.
(Note that the Berry curvature is antisymmetric under this exchange.)
The first term in the integrand is the contribution from the group velocity of the band.
Because of the $90^\circ$ rotational symmetry of the lattice and of the source, one can show that the symmetric term $\mathcal{A}_{\mathrm{sym}}(\mathbf{k})$ and the group velocity term should vanish upon integration over the momentum space.
(This is equivalent to noticing that the displacement in $x$ direction when the force $F$ is directed toward negative $y$ direction is the same as the displacement in $y$ direction when the force $F$ is directed toward positive $x$ direction.
Or mathematically, performing a transformation $\partial/\partial k_x \to \partial/\partial k_y$, $\partial/\partial k_y \to \partial/\partial k_x$, and $F \to -F$ should leave the expression invariant.)
Then, dividing (\ref{numresult}) by (\ref{denresult}), one obtains the desired relation Eq.~(3).

\setcounter{equation}{0}
\renewcommand{\theequation}{B\arabic{equation}}
\section{Appendix B: Details of the honeycomb lattice}

Here we give details of how to apply our scheme to photonic honeycomb lattices.
The geometry of the honeycomb lattice is shown in Fig.~3(a) of the main text.
We consider a honeycomb lattice with sublattice-dependent on-site energy described by the Hamiltonian
\begin{align}
	&H
	=
	\notag \\
	&
	\sum_{m,n}
	\left[
	-J \left(
	\hat{a}_{m+1,n}^\dagger \hat{b}_{m,n}
	+
	\hat{a}_{m,n+1}^\dagger \hat{b}_{m,n}
	+
	\hat{a}_{m,n}^\dagger \hat{b}_{m,n}
	+
	h.c.
	\right)
	\right.
	\notag \\
	&
	\left.
	+
	\left(
	\frac{\Delta}{2}
	+
	Fy
	\right)
	\hat{a}_{m,n}^\dagger \hat{a}_{m,n}
	+
	\left(
	-
	\frac{\Delta}{2}
	+
	Fy
	\right)
	\hat{b}_{m,n}^\dagger \hat{b}_{m,n}
	\right],
\end{align}
where $\hat{a}_{m,n}$ and $\hat{b}_{m,n}$ are the annihilation operators of photons in sublattices $A$ and $B$ with positions $m\mathbf{a}_1 + n\mathbf{a}_2$ and $m\mathbf{a}_1 + n\mathbf{a}_2 + (1,0)$, respectively.
The symbol $h.c.$ stands for the Hermitian conjugate.
As before, we assume that a weak external force $F$ is applied to the photons along the $-y$ direction; $y = \sqrt{3}(m-n)/2$ is indeed the $y$-coordinate of the $m,n$ site.

Because of the sublattice-dependent energy $\Delta \neq 0$, the dispersion is now gapped and the two bands have the energies
\begin{align}
	E_\pm (\mathbf{k})
	=
	\pm \sqrt{(\Delta/2)^2 + |g(\mathbf{k})|^2},
\end{align}
where $g(\mathbf{k}) \equiv -J \left[ e^{i\mathbf{k}\cdot\mathbf{a}_1} + e^{i\mathbf{k}\cdot\mathbf{a}_2} + 1 \right]e^{-ik_x}$.
The Dirac points where the separation between the bands is $\Delta$ are
\begin{align}
	\mathbf{K} &\equiv \frac{2\pi}{3}\left(1,\frac{1}{\sqrt{3}} \right),
	&
	\mathbf{K}^\prime &\equiv \frac{2\pi}{3}\left(1,-\frac{1}{\sqrt{3}} \right),
\end{align}
and those points which are obtained by translating $\mathbf{K}$ and $\mathbf{K}^\prime$ by reciprocal lattice vectors.
The Berry curvature, as shown in Fig.~3(b) for the upper band, has nonzero values around both Dirac points $\mathbf{K,K}^\prime$. Their contributions however cancel out when the curvature is integrated over the whole Brillouin zone to obtain the Chern number. The Berry curvature of the lower band has the same magnitude with the opposite sign.

We focus on the point $\mathbf{K}$ and consider the Berry curvature in the vicinity of this point.
The dispersion and the Berry curvature around this point has a simple form~\cite{Xiao2007S}; defining $\mathbf{q} \equiv \mathbf{k} - \mathbf{K}$, one has
\begin{align}
	E_{\pm}(\mathbf{q}) &\approx \pm \sqrt{(\Delta/2)^2 + (3Jq/2)^2},
	\notag \\
	\Omega (\mathbf{q})
	&\approx
	\mp
	\frac{9}{16}\frac{\Delta J^2}{\{\Delta^2/4+(3Jq/2)^2\}^{3/2}}. \label{omegaaroundK}
\end{align}
Since the energy and the Berry curvature are isotropic and either monotonically increasing or decreasing around $\mathbf{K}$ as a function of $|\mathbf{q}|$, the Berry curvature is a function of energy when the momentum is close to the Dirac point, which suggests that it is possible to measure the Berry curvature as a function of the frequency, $\Omega (\omega_0)$, using our proposal.

As we have mentioned in the main text, difficulties arise in the honeycomb case by the presence of two inequivalent sites within each unit cell, the so-called $A$ and $B$ sublattices.
A strategy to overcome this unwanted feature is to drive the system with a very wide spot of Gaussian shape (6) with $\sigma\gg 1$ and a carrier momentum tuned right at the $\mathbf{K}$ point. In this case, only a close neighborhood of the the tip of the Dirac cones is excited and the wavefunction on one of the sublattices become negligible, so the complication arising from having two inequivalent sublattices disappear. We can then determine the Berry curvature exactly at gapped Dirac points by measurements with only one frequency through the relation 
\begin{equation}
\langle x \rangle \approx \frac{F \Omega(\mathbf{K})}{\gamma} + \mathrm{constant~term~in~} F.
\end{equation}
The factor of 2 of difference compared to Eq.~(5) is
because the integrands in Eq.~(3) are now multiplied by the momentum distribution of the source, which is concentrated at $\sim$$\mathbf{K}$ and thereby dominates the integral.
With this method, taking e.g. $\sigma=50$, we obtain a value $\Omega \approx -47.7$ for the Berry curvature at the $\mathbf{K}$ point, to be compared with the exact value $\Omega = -50$. All other parameters are taken to have the same values as in Fig. 3 (c) and the pump frequency is tuned at a (single) value right at band edge $\omega_0/J = 0.15$. As the momentum distribution of the source field is narrowly focused at the Dirac points, this method is not able to provide accurate estimates of the Berry curvature away from the Dirac points.

To measure the Berry curvature in a larger neighborhood of the Dirac points, a viable strategy is the two-frequency strategy discussed in the main text. The idea is to perform two measurements with opposite frequencies and combine the results to compute the mean displacement and then to calculate the Berry curvature.
With the pump frequency of $\omega_0$ and the pump field $f(\mathbf{R})$ focused around the momentum $\mathbf{K}$ defined in the main text Eq.~(6), the steady-state equations of motion analogous to Eq.~(2) are
\begin{align}
	&\left(
	\omega_0 + i\gamma - F y - \Delta/2
	\right)
	a_{m,n}
	\notag \\
	&\hspace{1cm}+
	J
	\left(
	b_{m-1,n} + b_{m,n-1} + b_{m,n}
	\right)
	=
	f(\mathbf{R}^a_{m,n})
	\notag \\
	&\left(
	\omega_0 + i\gamma -F y + \Delta/2
	\right)
	b_{m,n}
	\notag \\
	&\hspace{1cm}+
	J
	\left(
	a_{m+1,n} + a_{m,n+1} + a_{m,n}
	\right)
	=
	f(\mathbf{R}^b_{m,n}), \label{grapheneseom}
\end{align}
where $\mathbf{R}^a_{m,n}$ and $\mathbf{R}^b_{m,n}$ are the positions of the sites $a_{m,n}$ and $b_{m,n}$, respectively.

On the other hand, one can show that the photon amplitude generated by a pump of opposite frequency $-\omega_0$ is, up to unimportant phases, equivalent to the configuration when the pump frequency is $\omega_0$ and only the sign of the source field on $B$ sublattice is exchanged, with $x$ and $y$ axes being flipped.
Therefore, denoting both $a_{m,n}$ and $b_{m,n}$ by a single notation $a_{\mathbf{R}}$, the mean displacement defined by Eq.~(7) in the main text can be equivalently written as
\begin{align}
	\langle x \rangle_-
	&=
	\frac{\sum_{\mathbf{R}}R_x (|a_\mathbf{R}|^2 + |\tilde{a}_\mathbf{R}|^2)}
	{\sum_{\mathbf{R}}(|a_\mathbf{R}|^2 + |\tilde{a}_\mathbf{R}|^2)}, \label{meanxtilde}
\end{align}
where $\tilde{a}_\mathbf{R}$ is the photon field produced by the same condition as $a_\mathbf{R}$ except for a negative sign in front of the pumping field for $B$ sublattice.

We now give an outline of the proof that the mean displacement $\langle x \rangle$ satisfies Eq.~(3).
As in Appendix A, we Fourier transform the photon fields by
\begin{align}
	a_{m,n}
	&=
	\sum_{\mathbf{k}} e^{i\mathbf{k}\cdot(m\mathbf{a}_1 + n\mathbf{a}_2)} c_{a}(\mathbf{k})
	\notag \\
	b_{m,n}
	&=
	\sum_{\mathbf{k}} e^{i\mathbf{k}\cdot(m\mathbf{a}_1 + n\mathbf{a}_2 + (1,0))} c_{b}(\mathbf{k}).
\end{align}
Defining a two-component vector $|c(\mathbf{k})\rangle \equiv (c_a (\mathbf{k}), c_b (\mathbf{k}))$, the steady-state equation of motion analogous to Eq.~(\ref{blocheom}) is
\begin{align}
	\left[
	\left(
	\omega_0 + i\gamma - iF\frac{\partial}{\partial k_y}
	\right)
	I_2
	-
	H_\mathbf{k}
	\right]
	|c(\mathbf{k})\rangle
	=
	|f (\mathbf{k})\rangle,
\end{align}
where 
$
	H_\mathbf{k}
	\equiv
	\begin{pmatrix}
	\Delta/2 & g^* (\mathbf{k}) \\
	g (\mathbf{k}) & -\Delta/2
	\end{pmatrix}
$
is the Hamiltonian of a honeycomb lattice in the momentum space in the sublattice basis, and $|f(\mathbf{k})\rangle$ is the Fourier transform of the source.
For the photon fields $a_\mathbf{R}$, $|f(\mathbf{k})\rangle$ takes the following form:
\begin{align}
	|f(\mathbf{k})\rangle
	=
	f(\mathbf{k})
	\begin{pmatrix}
	1 \\ 1
	\end{pmatrix}
	\equiv
	f(\mathbf{k}) |+\rangle. \label{sourcep}
\end{align}
On the other hand, for the photon fields $\tilde{a}_\mathbf{R}$, we have
\begin{align}
	|f(\mathbf{k})\rangle
	=
	f(\mathbf{k})
	\begin{pmatrix}
	1 \\ -1
	\end{pmatrix}
	\equiv
	f(\mathbf{k}) |-\rangle. \label{sourcem}
\end{align}
Then, Eq.~(\ref{meanxtilde}) becomes
\begin{align}
	\langle x \rangle_-
	=
	\frac{\sum_{k_x, k_y} \langle c(\mathbf{k}) | i\frac{\partial}{\partial k_x}| c(\mathbf{k})\rangle +\langle \tilde{c}(\mathbf{k}) | i\frac{\partial}{\partial k_x}| \tilde{c}(\mathbf{k})\rangle}
	{\sum_{k_x, k_y}\langle c(\mathbf{k}) | c(\mathbf{k})\rangle + \langle \tilde{c}(\mathbf{k}) | \tilde{c}(\mathbf{k})\rangle} \label{xfouriergraphene}
\end{align}
where $| c(\mathbf{k})\rangle$ is the Fourier transform of the photon fields when the source is (\ref{sourcep}), and $| \tilde{c}(\mathbf{k})\rangle$ is when the source is (\ref{sourcem}).
After a straightforward algebra, one can derive
\begin{align}
	\langle x\rangle_-
	=\frac{
	\int d^2 k
	f(\mathbf{k})
	\left\{
	\gamma \frac{\partial \mathcal{E}}{\partial k_x}
	+
	F
	\left(
	\gamma \Omega (\mathbf{k})
	+
	\mathcal{A}_{\mathrm{sym}}(\mathbf{k})
	\right)
	\right\}
	n(\mathbf{k})^2}
	{\int d^2 \mathbf{k}\,f(\mathbf{k}) n(\mathbf{k})}
\end{align}
which is analogous to (\ref{denresult}) and (\ref{numresult}) for the square lattice.
Upon derivation one uses the relation $|+\rangle \langle +| + |- \rangle \langle - | \propto I_2$, which is exactly the reason we take two measurements and combine them.
Similarly to the square lattice, because of the $120^\circ$ rotational symmetry of the lattice and the source, one can again show that the symmetric term and the group velocity term should vanish upon integration. Alternatively, this can be explicitly shown by expanding the relevant quantities in terms of small $\mathbf{q} = \mathbf{k} - \mathbf{K}$.
This directly leads to the desired relation Eq.~(3) with its integrand multiplied by the weight $f(\mathbf{k})$.